# Creation of a new vector field and focusing engineering

Xi-Lin WANG[1], Jing CHEN[1], Yongnan LI[1], Jianping DING[2], Cheng-Shan GUO[1], and Hui-Tian WANG[1,2,*]

[1]School of Physics & Key Laboratory of Weak Light Nonlinear Photonics, Nankai University, Tianjin 300071, China

[2]Nanjing National Laboratory of Microstructures, Nanjing University, Nanjing 210093, China

*e-mail: htwang@nankai.edu.cn or htwang@nju.edu.cn

**Recently many methods have been proposed to create the vector fields, due to the academic interest and a variety of attractive applications such as for particle acceleration, optical trapping, particle manipulation, and fluorescence imaging. For the most of the created vector fields, the spatial distribution in states of polarization (SoPs) is dependent of azimuthal angle only. It is very interesting and crucial that if we can introduce the radial controlling freedom, which undoubtedly opens a new way to provide the flexibility for creating the desired vector fields and for fulfilling the requirement on a variety of applications. Here we present a new idea to create a new kind of vector filed with the radial-variant SoPs. This idea also permits to create flexibly vector fields with arbitrarily complex distribution of SoPs, based on a combination of radial and azimuthal dependency. This realization in both principle and experiment is paramount to be able to implement the focusing engineering for applications in a variety of realms. Specially arranging the SoPs of vector fields, purposefully and carefully, is anticipated to lead to new effects and phenomena that can expand the functionality and enhance the capability of optical systems, such as the optical trapping.**

The purely transverse electric mode was firstly achieved in a Ruby laser by utilizing an intracavity axial birefringent component in 1972[1,2]. In the resent years radially polarized fields as a kind of typical vector fields have greatly developed and have extensively attracted attention[3-10]. This is mainly due to many novel attractive applications caused by some peculiar features, in various realms such as nonlinear optics, near-field optics, optical trapping and particle manipulation, imaging, spectroscopy, nanotechnology, micromechanics, microfluidics, and biology. The great progress in the creation of vector fields has opened a window for manipulating the states of polarization (SoPs) of light field and has provided the feasibility for implementing the desired applications. For instance, focusing by an objective with a high numerical-aperture (NA), the radially polarized fields as a typical kind of vector fields generate a strong longitudinal and non-propagating electric field at the focus[11-13]. With purposefully and carefully designed pupil mask, this longitudinal component could result in a sharper focal spot than a scalar field (including linearly, circularly and elliptically polarized)[11,14-16]. This distinct feature makes the radially polarized fields are better than other scalar fields for many applications, such as excitation of surface plasmon-polaritons[17-20], microscopy[21-23], near-field optics[24], fabrication of photonic crystals[25] and optical trapping and manipulation of metallic particles[26,27]. The azimuthally polarized fields, as another typical kind of vector fields, can be highly focused into a hollow dark spot[11-13] and used for focusing engineering[13,28-30]. Furthermore, focusing a vector filed with a combination of radially and azimuthally polarized fields could be highly focused into an optical plat[13,28] or a three-dimensional optical cage[29,30].

However, the creation of novel vector fields is still a big challenge, due to the high expectation of feasibility for flexibly manipulating the states of polarization (SoPs) of light field and then for giving full plays to the potential.

## CREATION AND RESULTS

Most of the vector fields reported in literature have a common feature that, as schematic drawing shown in Fig. 1a, the spatial distribution of SoPs is dependent of azimuthal angle $\phi$ only while is independent of the radial vector $\rho$. We here refer to such a kind of vector fields as the azimuthal-variant vector fields. The field distribution in the cross section of an azimuthal-variant vector field can be expressed by[10]

$$\vec{E}(\rho,\phi) = A(\rho)\left[\cos(m\phi - \phi + \alpha_0)\hat{e}_\rho + \sin(m\phi - \phi + \alpha_0)\hat{e}_\phi\right], \quad (1)$$

where $A(\rho)$ represents the radial-dependent amplitude, $\hat{e}_\rho$ and $\hat{e}_\phi$ are the unit vectors in the polar coordinate system ($\rho$, $\phi$), $m$ is the topological charge or the azimuthal index denoting the number of rotation cycles of SoP around the beam axis when the azimuthal angle $\phi$ rotates from 0 to $2\pi$, $\alpha_0$ decides the initial polarization direction at the field center of $\rho = 0$. When $m = 1$, $\alpha_0 = 0$ and $\pi/2$ correspond to the well-known radially and azimuthally polarized vector fields, respectively. It should be pointed out that equation (1) is valid when $m$ is a series of discrete integers; in this case the vector fields exhibit the cylindrical symmetry in intensity. When $m$ is an integer more than unity, the high-order azimuthal-variant vector fields[9,10,31] could be generated and could be tightly focused into a flower-like pattern[31]. It is very interesting that $m$ is a fraction, which makes polarization to be discontinuous along the ray direction of $\phi = 0$, resulting in the breaking of the cylindrical symmetry in intensity pattern, as reported in refs. 10 and 32. When $m$ is a fraction, the field distribution in the cross section of azimuthal-variant vector field can be still described by Eq. (1) within the range of $\phi \in (0, 2\pi)$, while in the ray direction of $\phi = 0$ (or $2\pi$) it should be expressed by

$$\vec{E}(\rho,\phi)\Big|_{\phi=0} = \frac{1}{2}A(\rho)\left\{\left[\cos(\alpha_0)\hat{e}_\rho + \sin(\alpha_0)\hat{e}_\phi\right] + \left[\cos(2m\pi + \alpha_0)\hat{e}_\rho + \sin(2m\pi + \alpha_0)\hat{e}_\phi\right]\right\}, \quad (2)$$

In particular, if *m* is a half-integer, we have $\vec{E}(\rho,\phi)|_{\phi=0} = 0$, implying that there is always a zero-intensity ray along the ray direction of $\phi = 0$. In contrast, if *m* is not a half-integer, we yield $0 < |\vec{E}(\rho,\phi)|_{\phi=0}| < A(\rho)$, suggesting that there is a dark ray along the ray direction of $\phi = 0$.

It is very interesting that could we create a kind of new vector field, as schematic drawing shown in Fig. 1b, in which the spatial distribution of SoPs is dependent of radial vector $\rho$ only while is independent of azimuthal angle $\phi$. This makes us remember the creation of the azimuthal-dependent vector fields in our previous work[10], in which the additional phase $\delta = m\phi + \alpha_0$ in the transmission function of spatial light modulator (SLM) is a function of azimuthal angle $\phi$ only. This gives for us a hint that it may be feasible that the additional phase $\delta$ becomes a function of the radial vector $\rho$ only, as $\delta = 2n\pi\rho/\rho_0 + \alpha_0$. We can yield the transverse field of a vector filed with its mathematic description, as follows

$$\vec{E}(\rho,\phi) = E_\rho \hat{e}_\rho + E_\phi e_\phi = A(\rho)\left[\cos\left(\frac{2n\pi\rho}{\rho_0} - \phi + \alpha_0\right)e_\rho + \sin\left(\frac{2n\pi\rho}{\rho_0} - \phi + \alpha_0\right)e_\phi\right]. \quad (3)$$

Here except for $\rho_0$ is the beam radius of the vector field and the real number *n* is referred as the radial index, the residual parameters have the same definitions as Eq. (1). Evidently, the spatial distribution of SoPs is a function of the radial vector $\rho$. It should be pointed out that when $n = 0$ the vector field described by equation (2) degenerates into a linearly polarized scalar field, like the case when $m = 0$ in equation (1).

The simulation results (are not shown here) validate that the vector fields described by the above equation are indeed dependent of radial vector $\rho$ only while independent of azimuthal angle $\phi$. We here can refer to such a kind of vector field as the radial-variant vector fields. We now demonstrate experimentally the feasibility of the radial-variant vector fields predicted in the above. We can utilize the same scheme in experiment as that in ref. 10 for creating the vector fields. The created four radial-variant vector fields at $\alpha_0 = 0$ are shown in Fig. 2a-d, corresponding to four different radial indices of $n = 0.5$, 1.0, 1.5, and 2.0. When no polarizer is used, the four vector fields have no difference and exhibit all the uniform distribution. In particular, these vector fields have an intrinsic difference from the azimuthal-variant ones that no singular spot occurs. It is easily understood that there is no local position of the polarization uncertainty, unlike the azimuthal-variant vector fields[10]. The intensity patterns of the radial-variant vector fields exhibit always the cylindrical symmetry for any radial index $n$, which is completely different from the azimuthal-variant vector fields, because the fractional azimuthal index $m$ gives rise to the cylindrical symmetry breaking[10]. After passing through a polarizer with its polarization direction along the horizontal direction, as shown in Fig. 2e-h, the intensity patterns exhibit the extinction rings, implying that the SoPs of the created vector fields are indeed radial-variant. The number of the extinction rings is nicely equal to $2n$ if $n$ is an integer or a half-integer. The radius of the $j$th extinction ring (where $j = 1, …, 2n$) is give by $\rho_j = (2j − 1)(\rho_0/4n)$. To intuitively show the spatial distribution of SoPs, the corresponding schematics of SoPs are drawn in Fig. 2i-l. SoP along the radial direction from $\rho = 0$ to $\rho_0$ is rotated by a degree of $2n\pi$ with respect to that at $\rho = 0$.

We now would like to explore the role of the parameter $\alpha_0$ describing the initial polarization direction at the field center of $\rho = 0$. As an example, the parameters $\alpha_0$ are chosen to be four values of $\alpha_0 = 0$, $\pi/4$, $\pi/2$, and $3\pi/4$, when the radial index $n$ is fixed to be 0.5. When no polarizer is used, the intensity patterns are not recognizable for all the four radial-variant vector beams with the different values of $\alpha_0$, as shown in Fig. 3a-d. When a polarizer is used, however, the intensity patterns become recognizable, and exhibit still the cylindrical symmetry with the homocentric extinction rings, as shown in Fig. 3e-h. The radii of the extinction rings are $\rho_0/2$, $\rho_0/4$, 0, and $3\rho_0/4$, for $\alpha_0 = 0$, $\pi/4$, $\pi/2$, and $3\pi/4$, respectively. Unlike the azimuthal-variant vector beams described by equation (1) have a central dark singular spot caused by the polarization uncertainty, the intensity patterns of four radial-variant vector beams have no singular spot at the center of $\rho = 0$, in which the local polarizations are determined by $\alpha_0$, as shown in Fig. 3i-l. For all the four radial-variant vector beams with $n = 0.5$, the polarization is rotated by a half of cycle as the radius increases from 0 to $\rho_0$. In detail, the local polarization at the center of $\rho = 0$ is along the direction of $\phi = \alpha_0$ and the polarization at any location with $\rho = \rho_0$ is at the direction of $\phi = \alpha_0 + \pi$.

It should be very attractive to create the simultaneous azimuthal- and radial-variant vector fields. Naturally, an easily conceivable idea is using a combination of two variables $\rho$ and $\phi$. We can choose the additional phase $\delta$ in the transmission function of SLM to have the form of $\delta = 2n\pi\rho/\rho_0 + m\phi + \alpha_0$, and then the transverse field of the created vector field can be described as follows

$$\overline{E}(\rho,\phi) = E_\rho \hat{e}_\rho + E_\phi e_\phi = A(\rho)\left[\cos\left(\frac{2n\pi\rho}{\rho_0} + m\phi - \phi + \alpha_0\right)e_\rho + \sin\left(\frac{2n\pi\rho}{\rho_0} + m\phi - \phi + \alpha_0\right)e_\phi\right] \quad (4)$$

The above equation degenerates to equation (1) describing the azimuthal-variant vector field when $n = 0$, while simplifies into equation (3) embodying the radial-variant vector field when $m = 0$.

Figure 4 furnishes the experimental results when setting $n = 1.0$ and $\alpha_0 = 0$, for four different values of $m$ = -1.0, 1.0, 2.0, and 3.0. Figure 5 shows the created vector fields when setting $m = 1.0$ and $\alpha_0 = 0$, for four different values of $n$ = -1.0, 2.0, and 3.0. From the schematics in Fig. 4a-d and Fig. 5a-c, we can find that the created vector fields have the more complex and rich SoPs. As shown in Fig. 4e-h and 5d-f, all the intensity patterns without the polarizer has a central singular spot caused by the polarization uncertainty when $m \neq 0$, like the azimuthal-variant vector fields while unlike the radial-variant vector fields. However, when a polarizer is inserted, as shown in Fig. 4i-l and Fig. 5g-i, the intensity patterns exhibit the Archimedean spiral structure, and the number of arms is nicely equal to $2|m|$ while is independent of *n*. The chirality of the Archimedean spiral is determined by the sign of the product of *m* and *n*. The chirality is right-handed screw if $mn > 0$ whereas it is left-handed screw if $mn < 0$. As we anticipated, the SoPs of the vector field described by equation (4) are indeed both azimuthal- and radial-variant spatial distribution.

Inasmuch as discretional radial index *n* does not influence on the cylindrical symmetry, we only need to explore the effect of the fractional azimuthal index *m* on the azimuthal- and radial-variant vector fields. In the experimental results of $m = 0.5$ (upper panel) and 1.5 (bottom panel) for setting $n = 1.0$ and $\alpha_0 = 0$, as shown in Fig. 6, any the intensity patterns contains a dark ray with its initial point located at the center along the direction of $\phi = 0$, which is the same as the azimuthal-variant vector fields. The dark ray caused by the fractional *m* breaks also the cylindrical symmetry for the azimuthal- and radial-variant vector field, which is the same as the azimuthal-variant one. It is clear that the schematics of SoPs as shown in Fig. 6a and Fig. 6d are markedly different and complex from that in the case of the integer *m*. The intensity patterns without a polarizer have still no difference, as Figs. 6b and 6e. However, if a polarizer is inserted, the intensity pattern exhibits still the Archimedean spirals (Figs. 6c and 6f) and the number of arms of the Archimedean spiral is always determined by $2|m|$ even though *m* is a half of integer. All the arms are smoothly connected at the center.

When the polarizer with its polarization along the direction of $\phi = 0$, we carefully deduce to give the Archimedean spiral equation as $\rho = a\phi + b$, where $a = -m\rho_0/2n\pi$ and $b = (\rho_0/2n\pi)[(2k-1)\pi/2 + \theta - \alpha_0]$. Here $\theta$ stands for the angle formed the polarization direction of the polarizer with the ray direction of $\phi = 0$, and *k* denotes the order of arms of the Archimedean spiral with $k = 1, 2, \ldots, 2|m|$. The parameter *b* determines the spiral turning, while the magnitude of *a* controls the distance between successive turnings. In addition, the sign of the parameter *a*, in fact the signs of both *m* and *n*, determines the chirality of the Archimedean spiral. In detail, the right-handed or left-handed screw corresponds to the case of $a < 0$ ($m$ and $n$ have the same signs) or $a > 0$ ($m$ and $n$ have the opposite signs). The extinction lines along the azimuthal direction are $2|m|$, while the extinction nodes along the radial direction are determined by $2|n|$.

## FOCUSING ENGINEERING AND OPTICAL TRAPPING

We would like to explore the focusing property of the radial-variant vector fields. When a radial-variant vector beam is tightly focused by a high numerical aperture (NA) objective, according to the Richards-Wolf vectorical diffraction theory, we can yield the electric field in the vicinity of focus, as $\vec{E}(r,\phi,z) = E_r \hat{e}_r + E_\varphi \hat{e}_\varphi + E_z \hat{e}_z$, where $\hat{e}_r$, $\hat{e}_\varphi$, and $\hat{e}_z$ are the unit vectors in the cylindrical system, with

$$
\begin{aligned}
E_r &= A \int_0^{NA} \left[ (\cos\beta + 1) J_0(kr\sin\beta) - (\cos\beta - 1) J_2(kr\sin\beta) \right] \\
&\quad \cos(2t\pi\sin\beta - \varphi) \cos^{1/2}\beta \sin\beta \exp(jkz\cos\beta) d\beta, \\
E_\varphi &= A \int_0^{NA} \left[ (\cos\beta + 1) J_0(kr\sin\beta) + (\cos\beta - 1) J_2(kr\sin\beta) \right] \\
&\quad \sin(2t\pi\sin\beta - \varphi) \cos^{1/2}\beta \sin\beta \exp(jkz\cos\beta) d\beta, \\
E_z &= 2jA \int_0^{NA} \sin\beta J_1(kr\sin\beta) \cos^{1/2}\beta \sin\beta \cos(2t\pi\sin\beta - \varphi) \\
&\quad \times \exp(jkz\cos\beta) d\beta
\end{aligned}
\quad (5)
$$

Here $A$ is a constant related to the focal length $f$ and the wavelength $\lambda$, and $J_0$, $J_1$ and $J_2$ are the Bessel functions of the first kind of orders 0, 1 and 2, respectively. The parameter $t$ is defined as $t = fn/\rho_0$, which is determined by the product of $f$ and $n$ for a given field radius $\rho_0$.

When implementing the numerical simulation for exploring the focusing property, we choose NA = 0.7 and $t = 9.72$, which are the same as the trapping experiments below. As the simulated intensity pattern in the focal plane in the inset of Fig. 7a, there are only a central bright spot with a size of $\sim\lambda$ and a bright ring with a radius $R = 9.486\lambda$ due to the constructive interference. Based on a series of simulations, the dependency of $R$ on $t$ is shown by the cross in Fig. 7a, which exhibits the near linear relationship. The linear fitting gives the linear relationship between $R$ and $t$, as shown by the solid line in Fig. 7a, as follows

$$R = (t - 0.234)\lambda \tag{6}$$

This formula suggests the focused bright ring to have the continuously changeable radius $R$, because $t$ is allowed to be any real number. The experimentally measured dependency of $R$ on $t$, as shown by the circle in Fig. 7a, is well agreement with the simulation results.

The ring focus could be used to trap microscopic particles. Since the first optical trap known as the optical tweezer was demonstrated[33], the optical tweezer has become a very useful tool for manipulating particles. A Gaussian laser beam is focused by a high NA objective, which can trap only a single particle. To generate the ring trap the scalar vortex field is in general required from an incident beam with a helical phase of $\exp(-jl\phi)$, as in refs. 34-37. Such a kind of ring focus has two characteristics. For the first one, the radius of the generated ring is only a series of discrete values while can never be continuously changed, as shown by the step line in Fig. 7b, because the topological charge $l$ must be an integer to ensure a ring focus. For the second one, due to the existence of helical phase, the particles trapped undergoes a torque and then will move along the ring focus[34,35] since each photon in a helical beam carries the orbital angular momentum of $l\hbar$[34,36]. A new type of ring trap formed by the radial-variant vector field has the continuously changeable radius and can satisfy the requirement that dispenses with the angular momentum.

We validate the capability of trapping particles by using the configurable ring trap generated by the radial-variant vector field we presented. The particles are polystyrene spheres with a diameter of 3.2 μm, which are dispersed in a layer of sodium dodecyl sulfate solution between a glass coverslip and a microscope slide. The laser used has a wavelength of 532 nm and a power of 20 mW. The incident radial-variant vector field has a radius of $\rho_0 =$ 2.5 mm. When a 60× objective with $f$ = 3.33 mm is used as a focal system to trap the particles, the experimental results are shown in Fig. 8. As mentioned above, in a given focal system and for a fixed radius of incident radial-dependent vector field, only the radial index $n$ can be used to change the ring radius $R$. Based on equation (6), when $n$ = 4.4, 5.8, 7.3, 8.8, 10.2 and 11.7, the radii of the generated focal rings are $R$ = 3.0, 4.0, 5.0, 6.0, 7.0 and 8.0 μm, correspondingly, the number of particles trapped in the focal ring is $N$ = 6, 8, 10, 12, 14, and 16, respectively, as shown in Fig. 8. It should be pointed out that a single particle is also trapped at the center of the focal ring, because there is a bright spot there. Due to the continuous tunability of the radius of the focal ring generated by the radial-variant vector field, such a kind of ring optical tweezer can trap arbitrary number of particles in the focal ring. If we can also introduce the orbital angular moment simultaneously, the functionality of such a kind of ring optical tweezer will become more powerful.

In conclusion, we have presented an idea to create a new kind of vector field, which is referred as the radial-variant vector field. Further, we can also create more complex vector field, which is both azimuthal- and radial-variant. The experiments have validated that based on our idea, the radial-variant vector filed and the both radial- and azimuthal-variant vector field could indeed be flexibly generated. Such new vector fields have some peculiar properties, with respect to the other optical fields. We have approved both theoretically and experimentally that the radial-variant vector field could be focused into a sharp ring with the continuously changeable radius and without the orbital angular moment, unlike the scalar optical vortex. The ring focus generated by the radial-variant vector field has successfully trapped the microscopic particles. The suggested method can be anticipated many useful applications.

This work is supported by the National Natural Science Foundation of China under Grant 10934003 and the National Basic Research Program of China under Grant 2006CB921805.

**FIGURE LEGENDS**

**Figure 1 Schematics of SoPs of two vector fields. a**, Azimuthal-variant vector field. **b**, Radial-variant vector field.

**Figure 2 Four radial-variant vector fields with $n$ = 0.5, 1.0, 1.5, and 2.0 when $\alpha_0$ = 0. a-d**, Respective intensity patterns without the polarizer. **e-h**, Respective intensity patterns with the polarizer. **i-l**, Respective schematics of SoPs.

**Figure 3 Four radial-variant vector fields with $\alpha_0$ = 0, $\pi/4$, $\pi/2$, and $3\pi/4$ when $n$ = 0.5. a-d**, Respective intensity patterns without the polarizer. **e-h**, Respective intensity patterns with the polarizer. **i-l**, Respective schematics of SoPs.

**Figure 4 Four simultaneously azimuthal- and radial- variant vector fields with $m$ = −1, 1, 2, and 3 when $n$ = 1.0. a-d**, Respective schematics of SoPs. **e-h**, Respective intensity patterns without the polarizer. **i-l**, Respective intensity patterns with the polarizer.

**Figure 5 Three simultaneously azimuthal- and radial- variant vector fields with $n$ = −1, 2, and 3 when $m$ = 1.0. a-c**, Respective schematics of SoPs. **d-f**, Respective intensity patterns without the polarizer. **g-i**, Respective intensity patterns with the polarizer.

**Figure 6 Two simultaneously azimuthal- and radial- variant vector fields with $m$ = 0.5 and 1.5 when $n$ = 1.0. a, d**, Respective schematics of SoPs. **b, e**, Respective intensity patterns without the polarizer. **c, f**, Respective intensity patterns with the polarizer.

**Figure 7 Properties of ring focus. a**, by radial-variant vector field, inset is a simulated ring focus. **b**, by scalar vortex field.

**Figure 8 Trapped particles in the ring focus. a-f**, corresponding to particle numbers of 6, 8, 10, 12, 14, and 16.

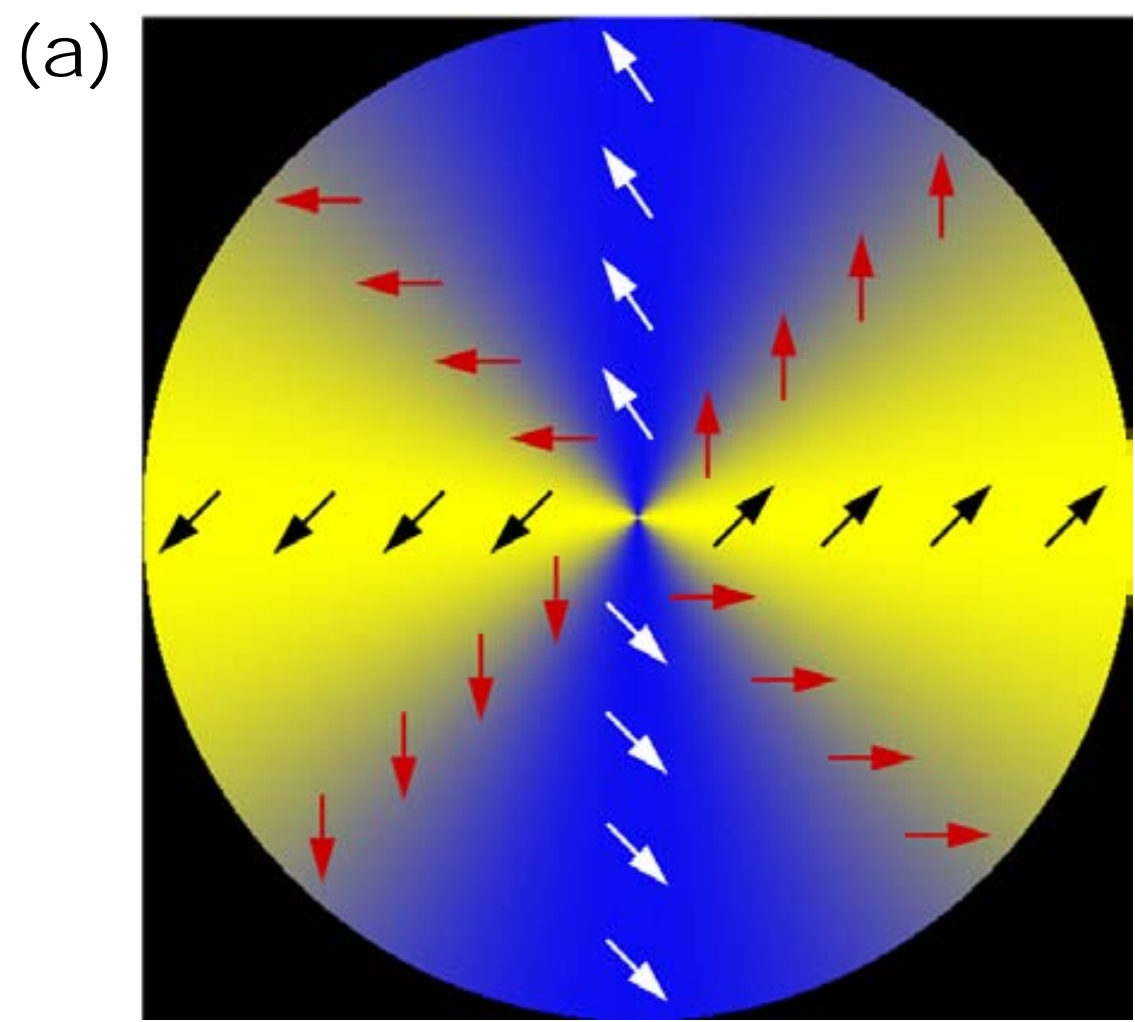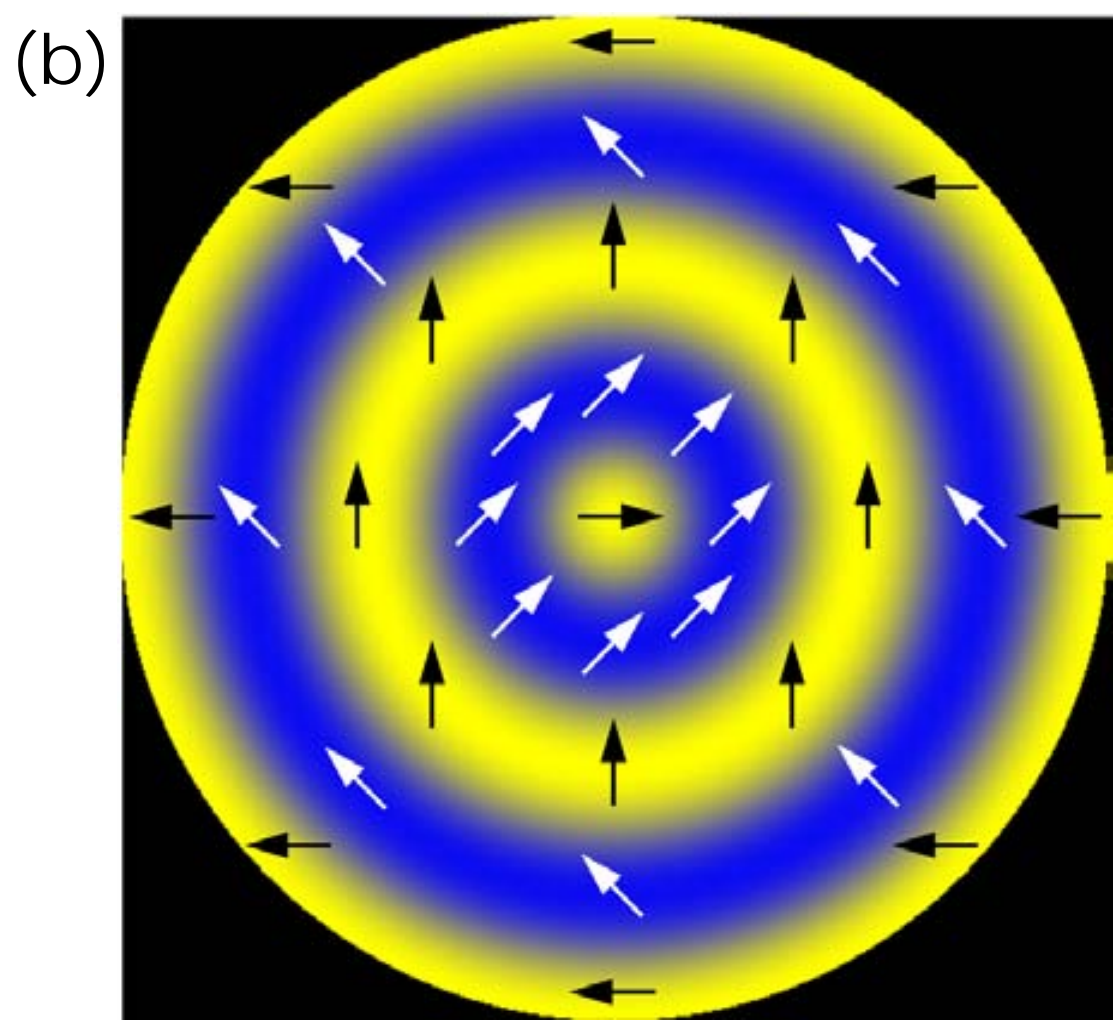

| n = 0.5 | n = 1.0 | n = 1.5 | n = 2.0 |
|---|---|---|---|
| 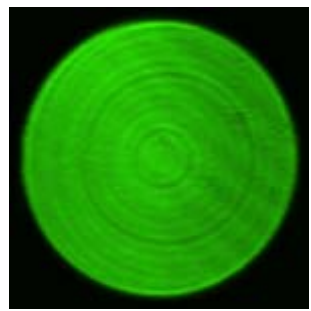 a | 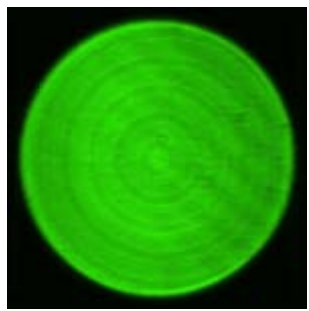 b | 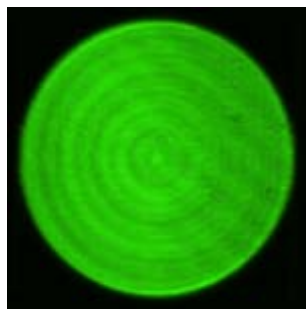 c | 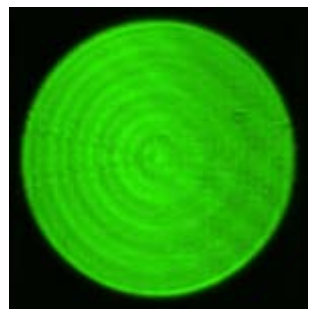 d |
| 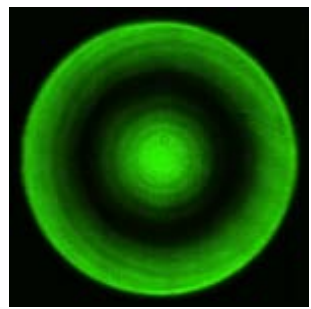 e | 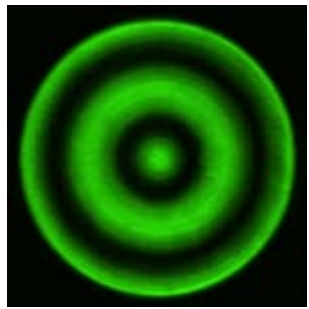 f | 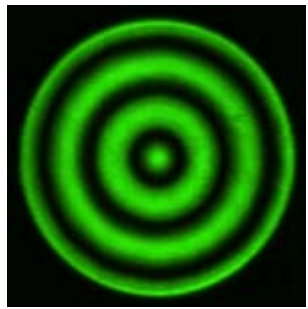 g | 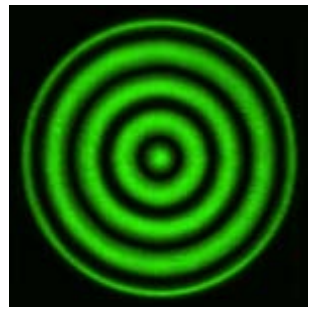 h |
| 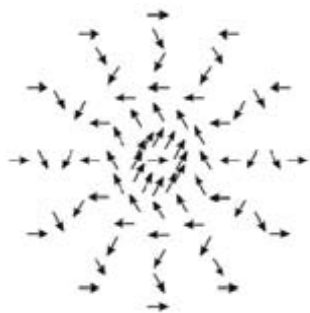 i | 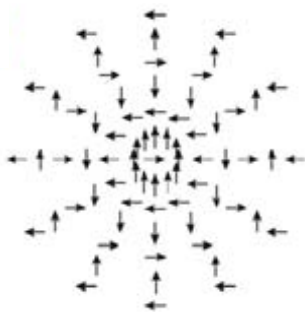 j | 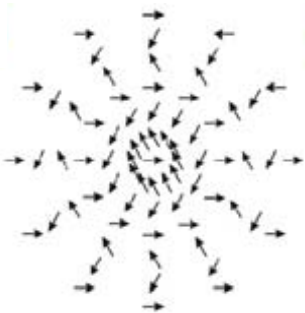 k | 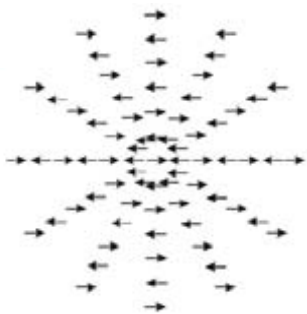 l |

| $\alpha_0 = 0$ | $\alpha_0 = \pi/4$ | $\alpha_0 = \pi/2$ | $\alpha_0 = 3\pi/4$ |

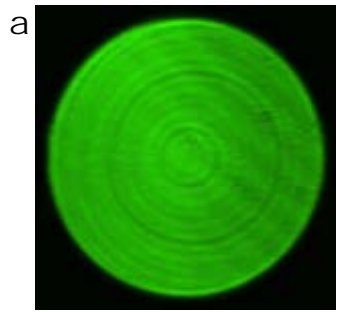 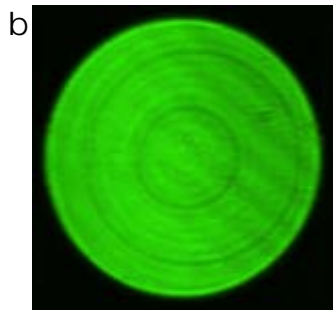 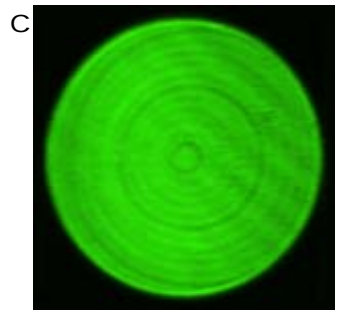 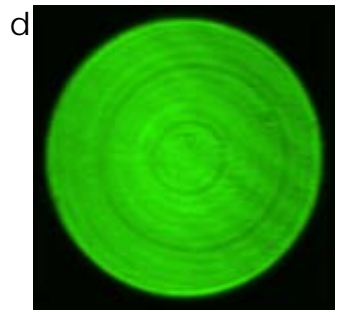

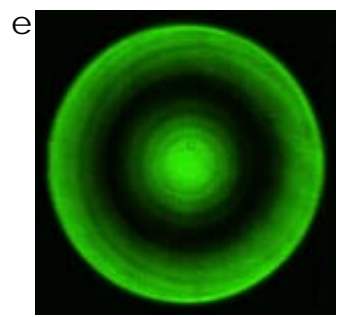 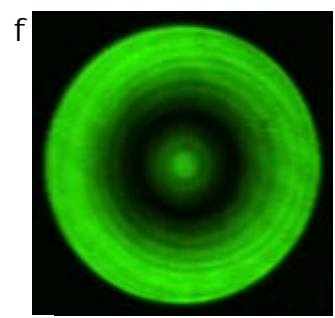 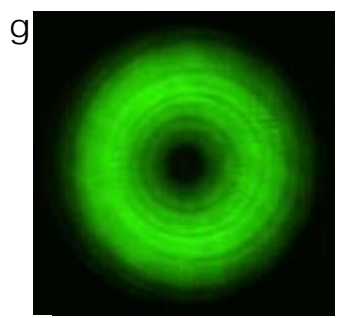 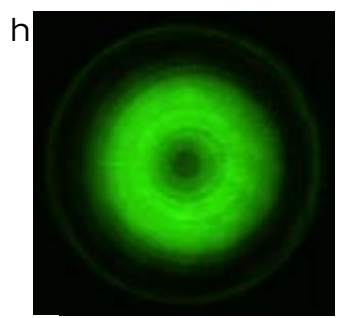

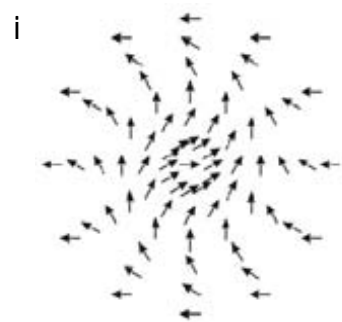 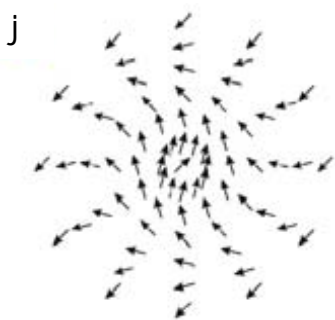 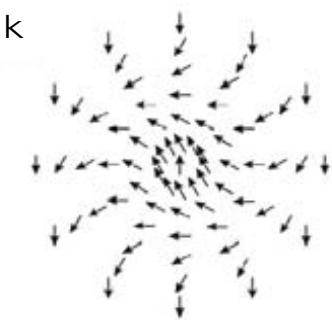 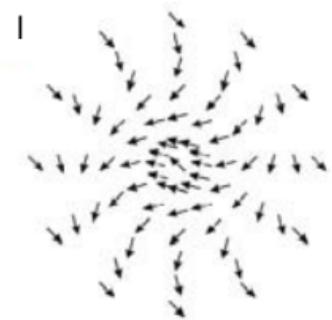

| m = −1 | m = 1 | m = 2 | m = 3 |
|---|---|---|---|
| 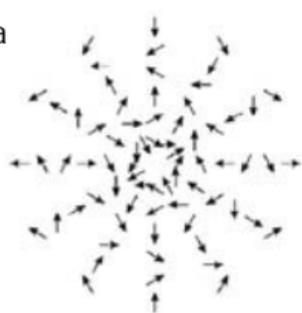 **a** | 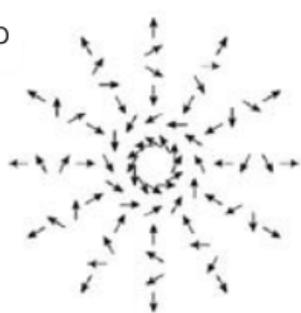 **b** | 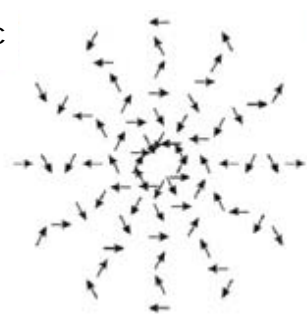 **c** | 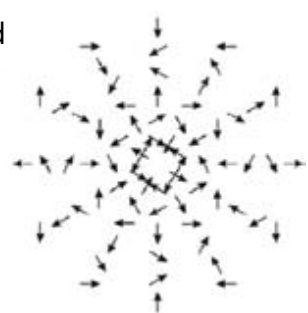 **d** |
| 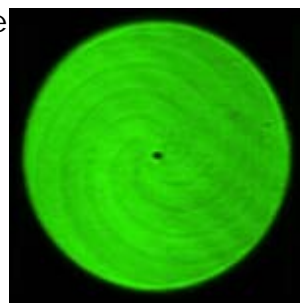 **e** | 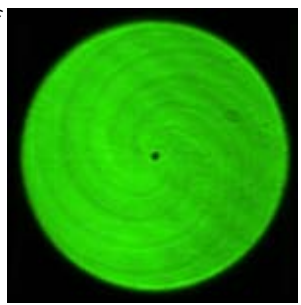 **f** | 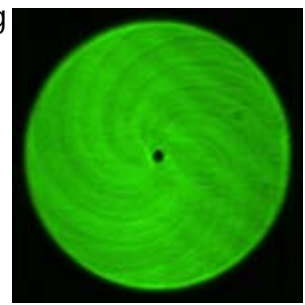 **g** | 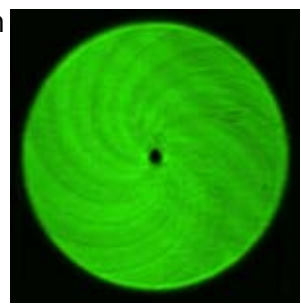 **h** |
| 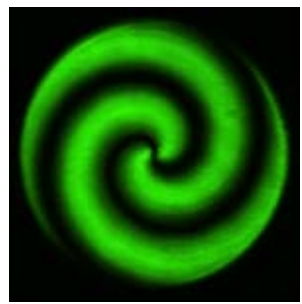 **i** | 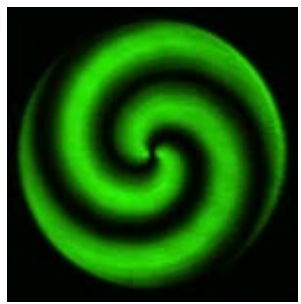 **j** | 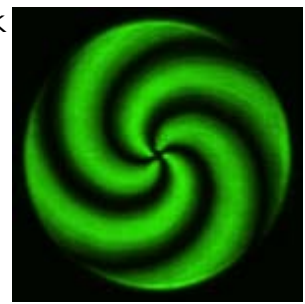 **k** | 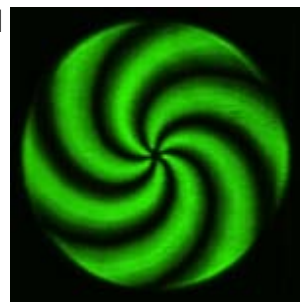 **l** |

| n = −1.0 | n = 2.0 | n = 3.0 |
|---|---|---|
| 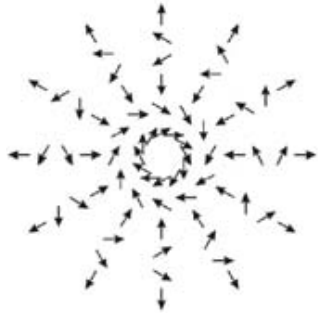 | 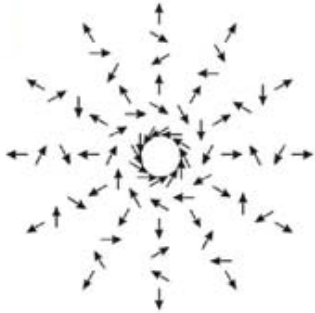 | 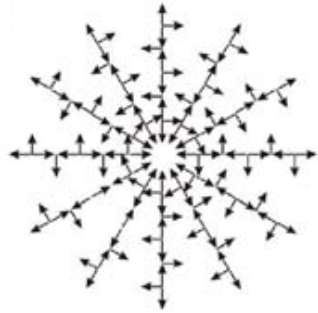 |
| a | b | c |
| 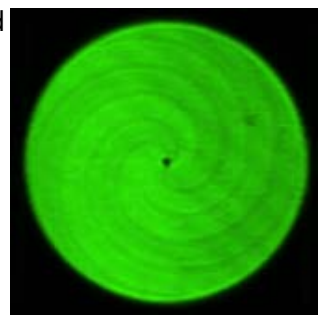 | 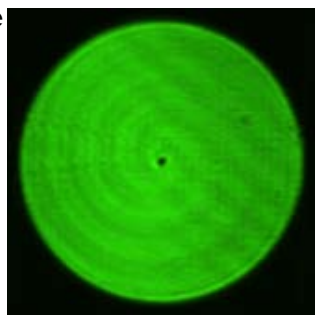 | 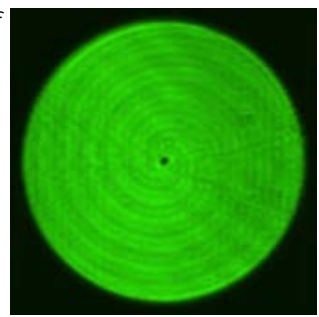 |
| d | e | f |
| 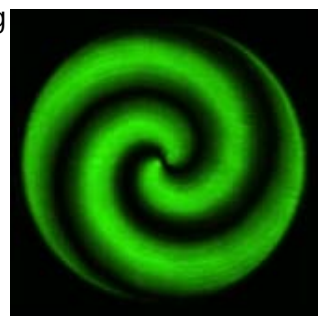 | 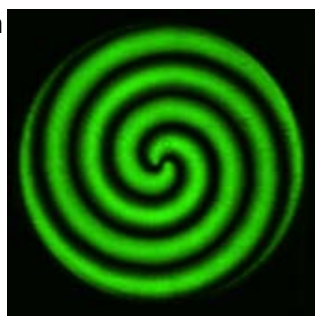 | 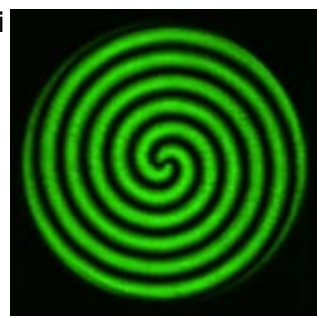 |
| g | h | i |

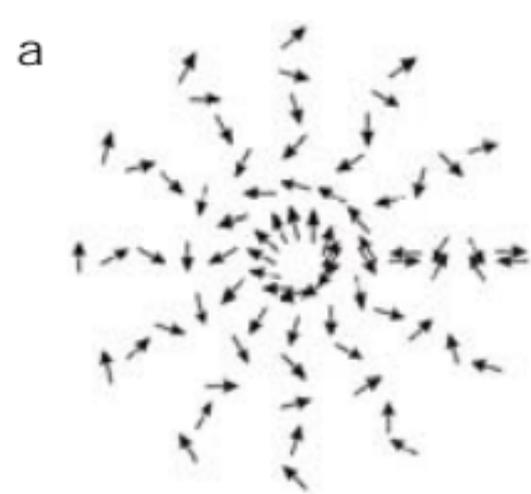 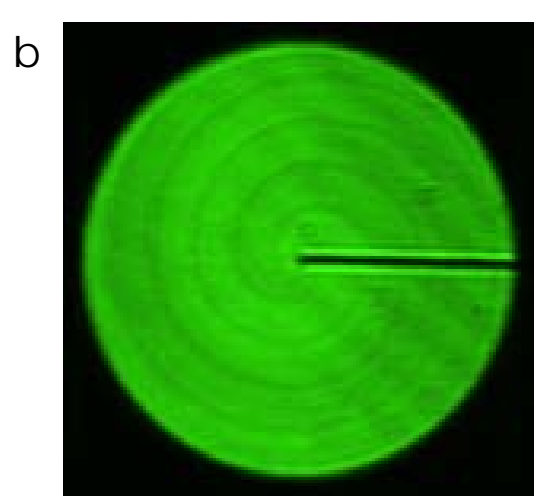 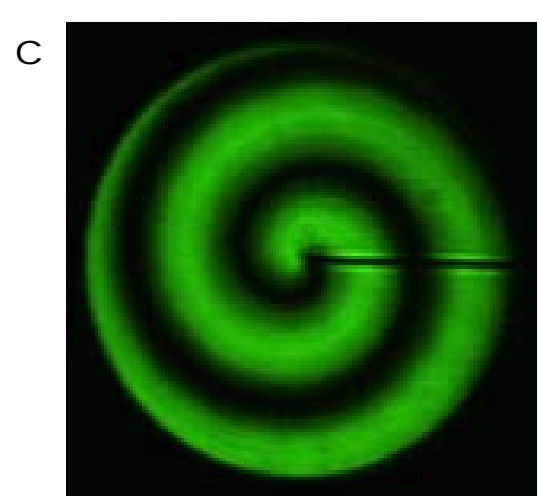
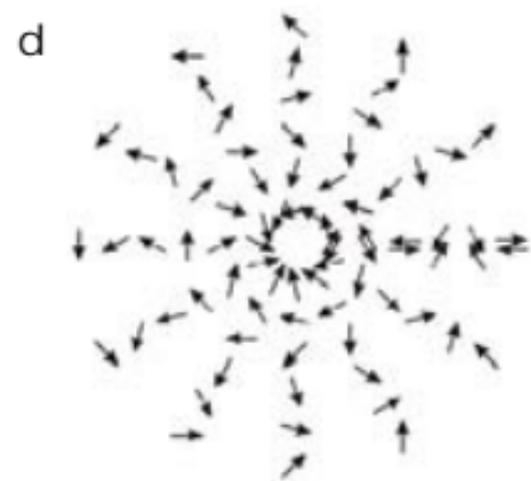 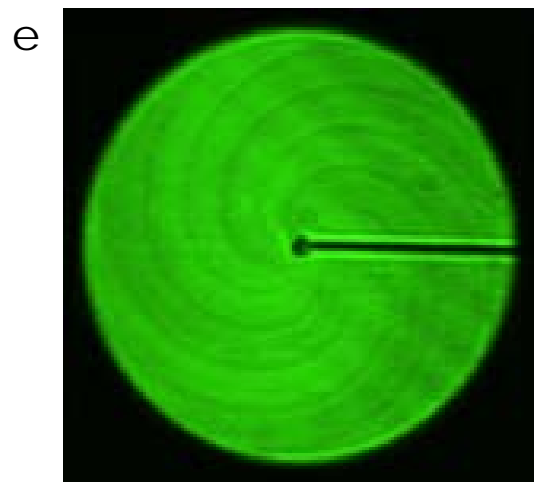 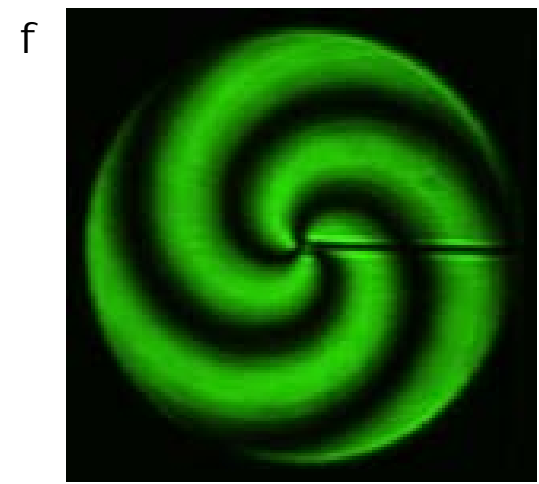

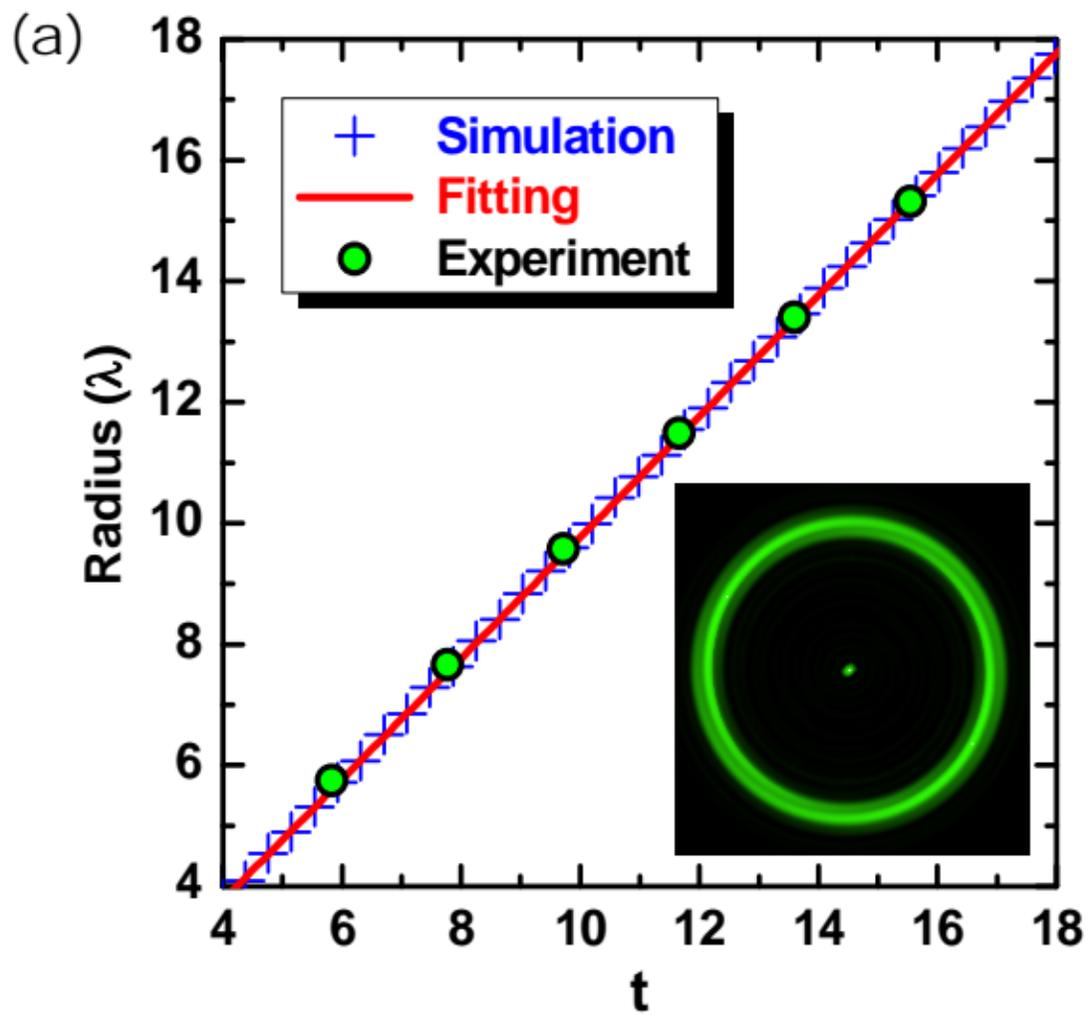 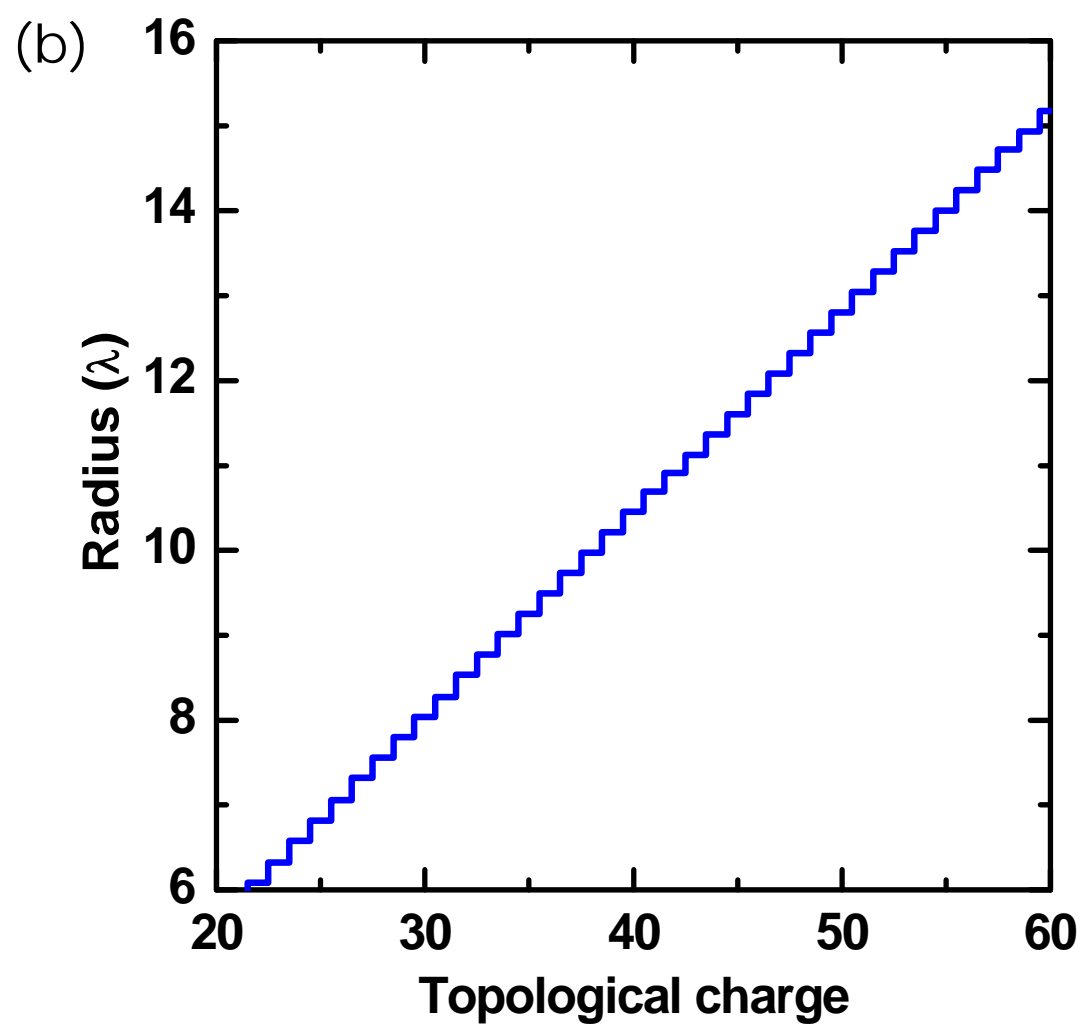

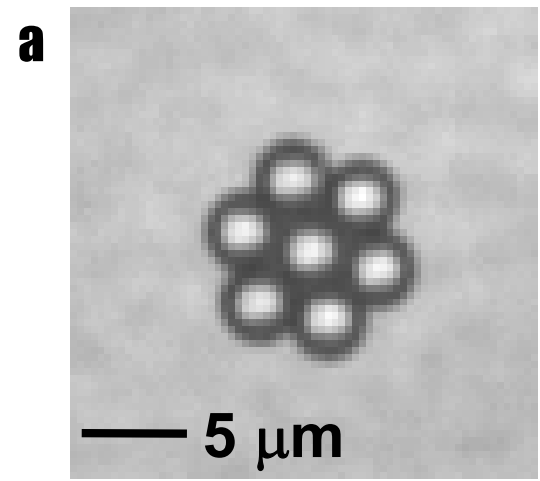 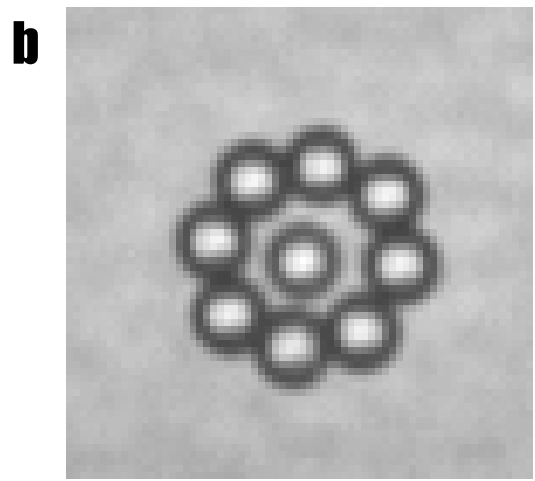 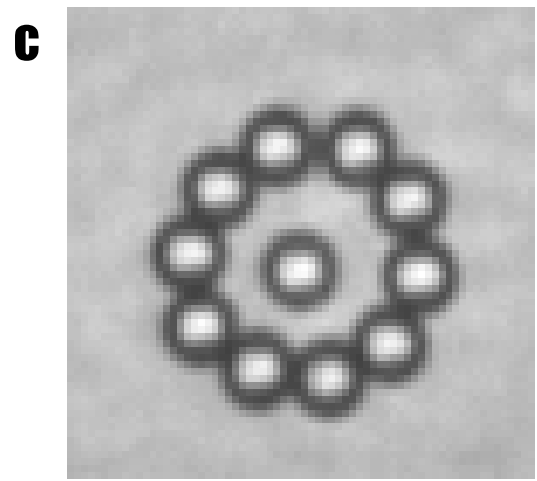
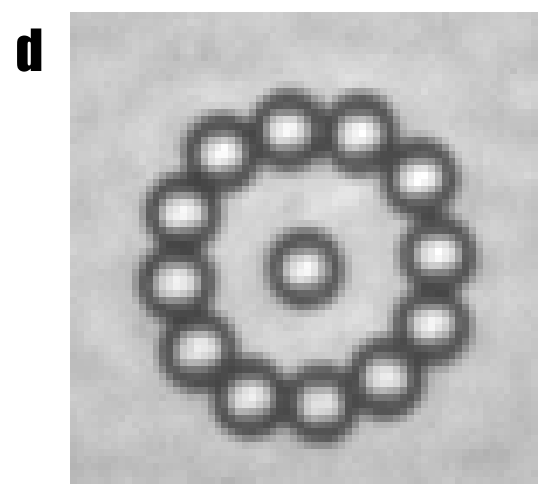 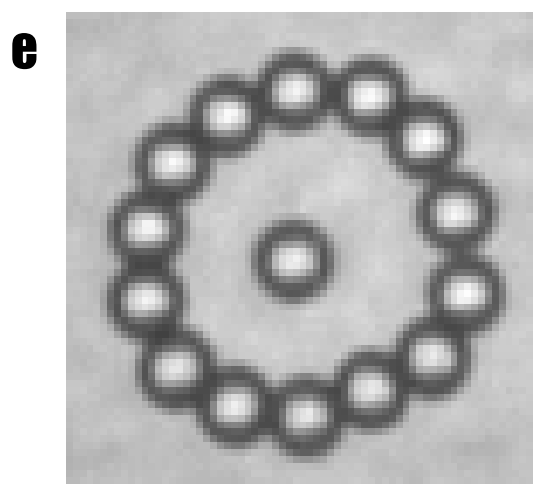 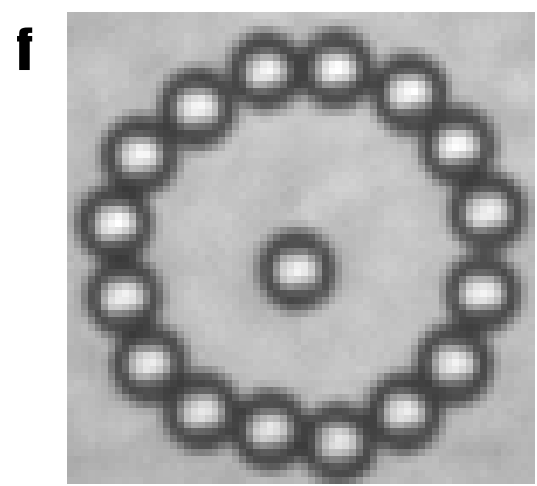

a  b  c
d  e  f

5 μm